\documentstyle[12pt,epsf]{article}
\newcommand{\resection}[1]{\setcounter{equation}{0}\section{#1}}

\begin{document}
\vspace*{4cm}
\begin{center}
  \begin{Large}
  \begin{bf}
Indirect Effects of New Resonances\\
 at Future Linear Colliders\\
  \end{bf}
  \end{Large}
\end{center}
  \vspace{5mm}
\begin{center}
  \begin{large}
R. Casalbuoni, S. De Curtis and D. Dominici\\
  \end{large}
Dipartimento di Fisica, Universit\`a di Firenze\\
I.N.F.N., Sezione di Firenze\\
\end{center}
  \vspace{5mm}
  \vspace{2cm}
\begin{center}
University of Florence - DFF 271/01/97
\end{center}
\vspace{1cm}
\noindent
\newpage
\thispagestyle{empty}
\begin{quotation}
\vspace*{5cm}

\begin{center}
\begin{bf}
  ABSTRACT
  \end{bf}
\end{center}
\vspace{1cm}
\noindent
In this paper we consider a general $SU(2)_L\otimes SU(2)_R$
invariant Lagrangian describing scalar, vector and
axial-vector resonances. By expanding the $WW$ and the $ZZ$
scattering amplitude up to the fourth order in the external
momenta we can compare the parameters of our Lagrangian with
the ones used in the effective chiral Lagrangian formalism.
In the last approach there has been a recent study of the
fusion processes at future $e^+e^-$ colliders at energies
above 1 $TeV$. We use these
results to put bounds on the parameter space of our
model and to show
that for the case of vector resonances the bounds obtained
from the annihilation channel in fermion pairs are by far
more restrictive, already at energies of the order of $500$
$GeV$.

  \vspace{5mm}
\noindent
\end{quotation}
\newpage
\setcounter{page}{1}

\newcommand{\be}{\begin{equation}}
\newcommand{\ee}{\end{equation}}
\newcommand{\bea}{\begin{eqnarray}}
\newcommand{\eea}{\end{eqnarray}}
\newcommand{\nn}{\nonumber}
\newcommand{\dd}{\displaystyle}
\newcommand{\gs}{{g''}^2}
\def\lmu{{\bf L}_\mu}
\def\rmu{{\bf R}_\mu}

\resection{Introduction}
 Future electron positron and proton
proton
colliders offer the possibility  of testing the nature of
spontaneous breaking of the electroweak symmetry,  the most
important open question  not yet clarified by the
high energy experiments done at LEP and Tevatron.

In particular if no light (supersymmetric) Higgs boson is
found, the $WW$ physics, accessible at such accelerators,
can explain which kind
of mechanism is responsible for the dynamical breaking of
the electroweak symmetry.
Different models of $WW$ scattering have been suggested, mainly
based on the chiral symmetry of the scalar sector of
the Standard Model (SM)
 and on the effective Lagrangian approach, both
with and without resonances. In fact if the $WW$ channel
becomes strongly interacting new resonances are expected, and
measuring the $WW$ amplitudes one can test the properties of these
new resonances.

A model describing  new vector, axial vector
resonances,
interacting with the standard gauge vector bosons
was proposed in
\cite{bess}\cite{assiali} (BESS model).
In this paper this model is further generalized
to include also a  scalar resonance.

The production of the vector resonances at LHC was studied
in a detailed way in a convenient region of the parameter space
where the production cross section is sizeable \cite{LHC}\cite{dege}.
Also we have studied the signature of this model at
linear electron positron
colliders  with  center of mass energy below the threshold
for the production  of these new resonances.
In this  low energy region
we have   studied indirect effects in the annihilation channel
both into fermions and $WW$ pairs
and in the fusion channel of the
$WW$ and $ZZ$ rescattering \cite{zeit}.

 In this paper the  $WW$ and $ZZ$ scattering amplitudes
are  obtained  by a general $SU(2)_L\times SU(2)_R$
invariant Lagrangian, which is derived by
an effective Lagrangian describing   scalar, vector, axial-vector
resonances, in the low energy limit up to fourth order in the
momentum expansion. This allows to find relations among
the parameters of our model and the ones commonly used in
the effective chiral Lagrangian formalism \cite{chiral}.

Concerning the limitations coming from future linear colliders,
our analysis  shows that the processes
$e^+ e^-\to W^+ W^- \bar \nu \nu$
and $e^+ e^-\to Z Z \bar \nu \nu$ are not the most important for testing
strongly interacting $WW$ system in presence of new vector resonances,
because the bounds on the relevant chiral parameters from the
annihilation channel are more stringent.

\resection{The model}

A model describing vector and axial-vector resonances
interacting in a consistent way with the three Goldstone bosons
necessary to give mass to the standard gauge vector bosons, $W$ and
$Z$, was considered in ref. \cite{assiali}. The idea is very
simple and consists in requiring that also the new resonances
acquire mass through the Higgs mechanism. In order to do that we
need to introduce at least nine Goldstone boson fields. The
other
requirement is that after spontaneous symmetry breaking the vector
resonances are triplets under the unbroken $SU(2)$. The simplest
way of realizing this situation is to assume beyond the standard
global symmetry $SU(2)_L\otimes SU(2)_R$, a local symmetry
$SU(2)$ for each new vector
resonance. This symmetry group $G\otimes H$ with

\be
G=[SU(2)_L\otimes SU(2)_R]_{\rm global},
~~~H=[SU(2)_L\otimes SU(2)_R]_{\rm local},
\ee
spontaneously breaks down to $SU(2)$.

The physics below the scale of the possible strong
interaction producing these resonances can be studied
 in terms of an effective Lagrangian. The methods to
construct such a Lagrangian are the standard ones used to build
up non-linear realizations (see ref. \cite{ccwz}). We will
describe the Goldstone bosons by three independent $SU(2)$
elements: $L$, $R$ and $M$, whose transformation properties
with respect to $G\otimes H$  are the following
\bea
&&L'(x)= g_L L(x) h_L(x),~~~R'(x)= g_R R(x) h_R(x),\nn
\eea
\be
M'(x)= h_R(x)^\dagger M(x) h_L(x)
\ee
where $g_{L,R}\in G$ and $h_{L,R}\in H$. Besides the invariance
under $G\otimes H$, we will also require an invariance under the
following
discrete left-right transformation, denoted by $P$
\be
L\leftrightarrow R,~~~~~M\leftrightarrow M^\dagger
\ee
which ensures that the low-energy theory is parity conserving.

The vector and axial-vector resonances are introduced as linear
combinations of the gauge particles associated to the local group
$H$. The most general $G\otimes H\otimes P$ invariant Lagrangian
is given by \cite{assiali}
\be
{\cal L}_R={\cal L}_G+{\cal L}_{kin}
\label{1.4}
\ee
where
\be
{\cal L}_G=-\frac{v^2}{4}f(\lmu,\rmu)
\label{1.5}
\ee
with
\be
f(\lmu,\rmu)= a I_1 + b I_2 + c I_3 + d I_4
\label{effe}
\ee
\bea
I_1=tr[(V_0-V_1-V_2)^2],&&I_2=tr[(V_0+V_2)^2]\nn\\
I_3=tr[(V_0-V_2)^2],&&I_4=tr[V_1^2]
\eea
and
\bea
V_0^\mu&=&L^\dagger D^\mu L\nn\\
V_1^\mu&=&M^\dagger D^\mu M\nn\\
V_2^\mu&=&M^\dagger(R^\dagger D^\mu R)M
\eea
The parameters $a$, $b$, $c$, $d$ are not independent
and we will fix later a condition among them,
in such a way that
 $v^2=1/(\sqrt{2}G_F)$.
The covariant derivatives are defined by
\bea
D_\mu L&=&\partial_\mu L -L \lmu\nn\\
D_\mu R&=&\partial_\mu R -R \rmu\nn\\
D_\mu M&=&\partial_\mu M -M \lmu+\rmu M
\eea
where $\lmu$ and $\rmu$ are the gauge fields associated to
the local symmetry group $H$.
The quantities $V_i^\mu~~(i=0,1,2)$ are, by construction,
invariant under the global symmetry $G$ and covariant under
the gauge group $H$
\be
(V_i^\mu)'=h_L^\dagger V_i^\mu h_L
\ee
Their transformation properties under the parity
operation, $P$, are:
\be (V_0\pm V_2)\to \pm M(V_0\pm V_2)M^\dagger,~~~~V_1\to - MV_1
M^\dagger
\ee
Out of the $V_i^\mu$ one can build six independent quadratic
invariants, which reduce to the four $I_i$ listed above, when parity
is enforced.
The kinetic part will be written in the form
\be
{\cal L}_{kin}=\frac{1}{\gs} tr[F_{\mu\nu}({\bf L})]^2+
         \frac{1}{\gs}  tr[F_{\mu\nu}({\bf R})]^2
\ee
where $g''$ is the gauge coupling constant for the gauge fields
$\lmu$
and
$\rmu$,
\be
F_{\mu\nu}({\bf L})=\partial_\mu{\bf L}_\nu-\partial_\nu{\bf L}_\mu+
          [\lmu,{\bf L}_\nu]
\label{effemunu}
\ee
and analogously for $\rmu$.

 We will also consider a scalar field invariant
under the group $G\otimes H\otimes P$. We will not be interested
in the self-couplings of this field $S$, and we will consider only
interaction terms with the previous fields at most linear in $S$.
The relevant effective Lagrangian is then
\be
{\cal L}= \frac 1 2\partial_\mu S\partial^\mu S-\frac 1 2 m^2 S^2
-\frac {v\kappa} 2 S f(\lmu,\rmu)+{\cal L}_R+\cdots
\label{ltot}
\ee
We can observe that in the case of a Higgs particle one has
$\kappa=1$.

\resection{Low-energy limit}

We will discuss now the low-energy limit of the previous
Lagrangian, by keeping terms up to the fourth order in the
derivatives. Let us start with the scalar field part.
At the lowest order we neglect the kinetic term. Then the equation
of motion gives
\be
S=-\frac{v\kappa}{2m^2}f(\lmu,\rmu)
\ee
We see that $S$ is at least of the second order in the
derivatives,
therefore the kinetic term
does not contribute at the lowest order,
and substituting in eq. (\ref{ltot}) we get
\be
{\cal L}=
\frac {v^2\kappa^2}{8m^2} [f(\lmu,\rmu)]^2+{\cal L}_R+\cdots
\ee
To discuss the same limit for the vector and axial-vector resonances
is convenient to choose the gauge $R(x)=M(x)=1$. This can be reached
by the gauge transformation $h_R(x)=R^{-1}(x)$, $h_L(x)=M^{-1}(x)
R^{-1}(x)$. By defining
\be
\omega_\mu=L^\dagger\partial_\mu L
\ee
we get
\be
V_{0\mu}=\omega_\mu-\lmu,~~~V_{1\mu}=-\lmu+\rmu,~~~V_{2\mu}
=-\rmu
\ee
If we neglect again the kinetic term for $\lmu$ and $\rmu$, we
can solve easily the equations of motion for the fields at the
leading order  in the derivatives.
It is convenient to put
\be
z=\frac c{c+d}
\ee
Then
\be
\rmu=\frac 1 2(1-z)\omega+\rmu^{(3)},~~~\lmu=\frac 1
2(1+z)\omega+\lmu^{(3)}
\label{soluzioni}
\ee
where $\rmu^{(3)}$ and $\lmu^{(3)}$ are of the third order in
the derivatives.
By substituting into $f(\lmu,\rmu)$ defined in (\ref {effe}) we find
\be
f(\lmu,\rmu)=(a+\frac{cd}{c+d})Tr(\omega^2)+\cdots
\ee
where the dots denote terms which are at least of the sixth order
in the  derivatives.
In order to recover the non-linear
$\sigma$-model describing the standard breaking
$SU(2)_L\otimes SU(2)_R$ to $SU(2)$  we  require a relation among
the parameters of the
Lagrangian (\ref{1.4}), namely
\be
a+\frac{cd}{c+d}=1
\ee
By using the following property, which holds for $2\times 2$
matrices of the form $\vec A\cdot\vec\sigma$, with $\vec\sigma$
the Pauli matrices,
\bea
Tr(ABCD)&=&Tr(AB)Tr(CD)-Tr(AC)Tr(BD)\nn\\
        &+&Tr(AD)Tr(BC)
\eea
we get
\be
(Tr(\omega^2))^2=Tr(\omega_\mu\omega^\mu\omega_\nu\omega^\nu)
\ee
and therefore
\be
f^2(\lmu,\rmu)= Tr(\omega_\mu\omega^\mu\omega_\nu\omega^\nu)
\ee
From the observation that $\omega_\mu$ is a 1-form with zero
curvature, that is satisfying
\be
 \partial_\mu\omega_\nu -\partial_\nu\omega_\mu+[\omega_\mu,
\omega_\nu]=0
\ee
and using
eq. (\ref{soluzioni}) and eq. (\ref{effemunu}), we get
\be
F_{\mu\nu}(\lmu)=F_{\mu\nu}(\rmu)=-\frac 1 4
(1-z^2)[\omega_\mu,\omega_\nu] +\cdots
\ee
Finally we find
\bea
{\cal L}&=&-\frac {v^2}4 Tr(\omega^2)+\frac 1 {4\gs}(1-z^2)^2
Tr[\,\omega_\mu\omega_\nu\omega^\mu\omega^\nu\,]\nn\\
&&+\left(
\frac{v^2\kappa^2}{8m^2}-\frac 1 {4\gs}(1-z^2)^2\right)
Tr[\,\omega_\mu\omega^\mu\omega_\nu\omega^\nu\,]+\cdots
\eea
This expression can be compared with the general expression for
the ungauged chiral Lagrangian \cite{chiral} at the same order
in the derivatives (we will do our calculations by using the
equivalence theorem \cite{equ}). We see that at the fourth order,
only the terms ${\cal L}_4$ and ${\cal L}_5$ are different from
zero (we are using the standard notations of ref.
\cite{chiral}), therefore we get the following expressions for
the chiral
parameters $\alpha_4$ and $\alpha_5$
\be
\alpha_4=\frac{(1-z^2)^2}{4\gs},~~~~~~
\alpha_4+\alpha_5=\frac{\kappa v^2}{8m^2}
\label{parameters}
\ee

Let us notice that this result holds
 for the model given by the Lagrangian
(\ref{ltot})
 where we have explicitly considered only terms with at most
two derivatives. Higher order terms in the derivative expansion add
contributions and the relations (\ref{parameters}) change.

\resection{Pion-Pion scattering amplitude}

In this Section we will study the scattering amplitude for the
$W^+_LW^-_L\to W^+_LW^-_L$ and $W_L^+W_L^-\to Z_LZ_L$, in
the case of the Lagrangian considered before, i.e. for the
case of a model with vector, axial vector and scalar particles.
We use the unitary gauge for the
$\lmu$ and $\rmu$ fields, characterized by the following
expressions for fields $L,~R,~M$:
\be
L=\exp(\frac{i\pi} v)\exp(\frac i v (\lambda-z\pi))
\ee
\be
R=\exp(\frac{-i\pi} v)\exp(\frac i v (\rho+z\pi))
\ee
\be
M= \exp(-\frac i v (\rho+z\pi))\exp(\frac i v (\lambda-z\pi))
\ee
where
\be
\pi=\vec\pi\cdot\frac{\vec\sigma} 2,~~~~\lambda=\vec\lambda\cdot
\frac{\vec\sigma} 2,~~~~\rho=\vec\rho\cdot
\frac{\vec\sigma} 2
\ee
It is not difficult to show that (apart from normalization
constants) $\vec\lambda$ and $\vec\rho$ are the Goldstone bosons
associated to $\lmu$ and $\rmu$ respectively, whereas $\vec\pi$
are the Goldstone bosons necessary to give mass to $W$ and $Z$.
The unitary gauge is then defined by taking
$\vec\lambda=\vec\rho=0$ in the previous expressions, that is
\be
L=\exp(\frac i v(1-z)\pi)=R^\dagger,~~~~M=\exp(-2\frac i v z\pi)
\ee
Using this gauge it is easy to evaluate the mass eigenstates
which are
${\bf V}_\mu=(\rmu+\lmu)/2$ and ${\bf A}_\mu=(\rmu-\lmu)/2$
with masses given respectively by
\be
M_V^2=b\frac {v^2} 4 \gs,~~~~M_A^2=(c+d)\frac {v^2} 4 \gs
\ee
In this gauge, from the $\pi\pi$ scattering amplitudes, by
using the equivalence theorem, we get
\be
M(W_L^+W_L^-\to Z_LZ_L)=A(s,t,u)
\label{zz}
\ee
\be
M(W_L^+W_L^-\to W_L^+W_L^-)=A(s,t,u)+A(t,s,u)
\label{ww}
\ee
where
\be
A(s,t,u)=\left(1-\frac 3 4 \beta\right)\frac s{v^2}-
\frac 1 4 \frac{M_V^2}{v^2}\beta\left(\frac{s-u}{t-M_V^2}+
\frac{s-t}{u-M_V^2}\right)-\frac{\kappa^2}{v^2}\frac{s^2}{s-m^2}
\label{amplitude}
\ee
and
\be
\beta= 4\frac{M_V^2}{\gs v^2}(1-z^2)^2
\ee
We can now expand the amplitude (\ref{amplitude})
 up to the fourth power in the  momenta
getting the following expression
\be
A(s,t,u)=\frac s{v^2}+\frac{(1-z^2)^2}{\gs
v^4}\left(-s^2+2st+2t^2\right)
+\frac{\kappa^2}{m^2 v^2}s^2
\label{2.4}
\ee
Comparing with the expression obtained by Kilian
\cite{kilian} from
the chiral Lagrangian \cite{chiral}
\bea
A(s,t,u)&=&\frac s{v^2}+4\alpha_4\frac{t^2+u^2}{v^4}+
8\alpha_5\frac{s^2}{v^4}\nn\\
&=& \frac s{v^2}+4\alpha_4\frac{-s^2+2st+2t^2}
{v^4}+
8(\alpha_4+\alpha_5)\frac{s^2}{v^4}
\eea
we get again the relation (\ref{parameters}) correlating the
chiral parameters $\alpha_4$ and $\alpha_5$ with the ones of the
model presented here.

Before considering some particular cases, let us discuss the
limitations on $\alpha_4$ and $\alpha_5$ coming from the
partial wave  unitarity conditions on the amplitudes (\ref{zz}),
 (\ref{ww}).
The most restrictive bounds come from $s$-wave. By requiring
$|a_0|\leq 1$, for $\sqrt{s}=1.6~TeV$,
we get the allowed region delimited from
the solid lines showed in Fig. 1.

\begin{figure}
\epsfysize=8truecm
\centerline{\epsffile{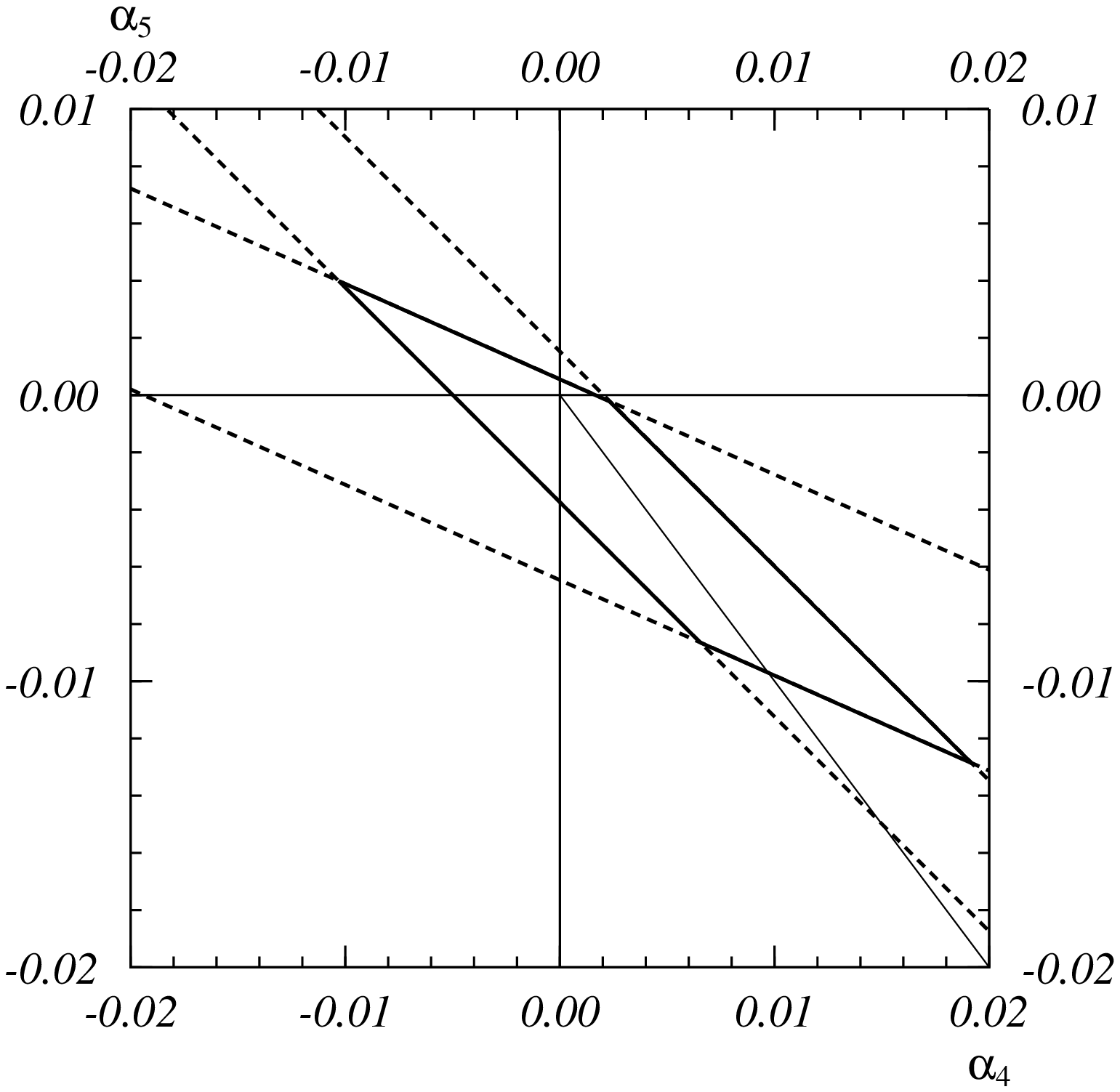}}
\noindent
{\bf Fig. 1} - {\it {Unitarity bounds on the plane
$(\alpha_4,\alpha_5)$ from
the $W_LW_L,Z_LZ_L\to W_LW_L$ scattering evaluated at
$\sqrt{s}=1.6~TeV$.
The allowed domain  is the diamond shaped region. The BESS model
corresponds to the line $\alpha_4+\alpha_5=0$ with $\alpha_4>0$.}}
\end{figure}
It is interesting to consider some particular case. We start
considering the vector case (BESS model \cite{bess}),
because here we can compare with
the analysis done at the $e^+e^-$ colliders.
We recall  that the restrictions on the
parameter space coming from the annihilation process
$e^+e^-\to W^+W^-$, have been already considered
in ref. \cite{zeit}. Therefore we will be able to
compare (at least for vector resonances) the sensitivity of the
two processes. For the BESS case there are no axial-vector and
scalar particles, that is we have to put $z=\kappa=0$.
We get
\be
\alpha_4=\frac{1}{4\gs},~~~~~~
\alpha_4+\alpha_5=0
\ee
This means that the BESS model lies on the line $\alpha_5=-\alpha_4$
with $\alpha_4\geq 0$ and from Fig. 1 we can see that the unitarity bound
is approximately $\alpha_4\leq 0.01$ which corresponds to $g''\geq 5$
(this is for $\sqrt{s}=1.6~TeV$).

From the Kilian \cite{kilian} analysis made for a
collider with  polarized
beams,
$\sqrt{s}=1600~GeV$ and an integrated luminosity of
$200~{\rm fb^{-1}}$ we find (for $\chi^2=3$, that is a confidence
level of 91.7\% for a gaussian distribution)
\be
\frac 1{4\gs}\le 0.0028
\ee
that is
\be
g''\ge 9.45
\label{1.11}
\ee
or
\be
\frac g{g''}\le 0.07
\ee
Following the analysis in ref. \cite{zeit}, for  a collider with
unpolarized beams,
$\sqrt{s}=500~GeV$ and an integrated luminosity of
$20~{\rm fb^{-1}}$ we find (again for $\chi^2=3$)
\be
\frac g{g''}\le  0.038
\ee
We see that the indirect bounds we get on the parameter space
from the annihilation process are by far more restrictive than
the ones for the fusion. The situation is, of course, much
different in the scalar channel, given the fact that this does
not contribute to the annihilation
process in a  collider $e^+e^-$.

The case of vector and axial-vector resonances is not too different,
but in this case, from the analysis of ref. \cite{kilian} one gets
bounds only for the combination $g''/(1-z^2)$. The bound will be the
same as given in (\ref{1.11}), that is
\be
\frac{g''}{1-z^2}\ge 9.45
\ee
An interesting observation is that the low-energy data give
informations on a different combination of $g''$ and $z$. In
fact the low-energy deviations from the SM, in the actual case,
can be entirely expressed  in terms of the observable $\epsilon_3$
 \cite{altarelli},
and this is given by \cite{selfenergies}
\be
\epsilon_3=\left(\frac{g}{g''}\right)^2\left(1-z^2\right)
\ee
Notice that in the case $z=1$, the so called
degenerate BESS model (vector and axial-vector resonances
degenerate in mass and in their couplings to the corresponding
currents) \cite{dege}, we loose any bounds;
in fact in the
scattering amplitude only the term corresponding to the
low-energy theorems \cite{LET} survives. This follows from the
fact that in the degenerate BESS the vector and axial-vector
resonances
decouple from the Goldstone bosons.

Finally for the case in which only the scalar particle is
present (or one has also vector and axial-vector resonances
degenerate), one
gets from \cite{kilian}
a bound on the combination $\kappa^2 v^2/m^2$ ($\kappa>0$)
\be
\frac{\kappa^2 v^2}{8m^2}\le +0.0012
\ee
suggesting that for a coupling of order 1, one can test up to
scalar masses of order of $2.55~TeV$.

\resection{Conclusions}
The BESS model provides an effective parametrization for strongly
interacting $WW$ system in presence of new resonances.
Below the resonance threshold, the relevant scattering amplitudes
within the BESS model can be compared with the results from
effective chiral Lagrangians.
Our analysis shows that
the fusion processes at future $e^+ e^-$ linear colliders
are not the most important for testing
strongly interacting $WW$ system in presence of new vector resonances.
In fact  the bounds on the BESS model parameters from the
annihilation channel, which we have studied in several previous papers,
are most stringent and can be already
obtained from colliders with a center of mass energy of 500 $GeV$.

\end{document}